# A granocentric model captures the statistical properties of monodisperse random packings

**Katherine A. Newhall, Ivane Jorjadze, Eric Vanden-Eijnden and Jasna Brujic**

We present a generalization of the granocentric model proposed in [Clusel *et al., Nature,* 2009, **460,** 611615] that is capable of describing the local fluctuations inside not only polydisperse but also monodisperse packings of spheres. This minimal model does not take into account the relative particle positions, yet it captures positional disorder through local stochastic processes sampled by efficient Monte Carlo methods. The disorder is characterized by the distributions of local parameters, such as the number of neighbors and contacts, filled solid angle around a central particle and the cell volumes. The model predictions are in good agreement with our experimental data on monodisperse random close packings of PMMA particles. Moreover, the model can be used to predict the distributions of local fluctuations in any packing, as long as the average number of neighbors, contacts and the packing fraction are known. These distributions give a microscopic foundation to the statistical mechanics framework for jammed matter and allow us to calculate thermodynamic quantities such as the compactivity in the phase space of possible jammed configurations.

## 1 Introduction and Model

The study of random packings of particles has received much attention in recent years. Its interest lies in uncovering possible packing geometries[1,2], understanding the local and global properties of granular materials[3] and glasses[4], and solving practical problems[5]. The diversity of theoretical approaches to packing therefore spans from geometric modeling[6] to analogies with glasses4 and the statistical mechanics of jammed matter[3]. They seek to quantify randomness in packed particle configurations through the distributions of local parameters, such as the coordination number, neighbor number and the cell volume of individual particles in a given packing. Experiments have measured these distributions as a function of the packing protocol for different particulate materials and thus tested theoretical approaches[7–14]. Nevertheless, a microscopic origin for the statistical fluctuations in random packings that bears out in experiments is still lacking in the literature[15].

While some look to understand packing from a macroscopic "thermodynamic" viewpoint [3,16–18] others turn to local microscopic descriptions[6,14,19,20]. In a recent work[19], we have developed a 'granocentric' model that is able to capture the geometric fluctuations inside a jammed polydisperse packing of spheres using the particle size distribution as the dominant source of randomness. This granocentric model is based on geometry alone and has also been shown to describe an unjammed system of disks[21] and may even be applicable to the random structure of liquids[22,23]. For jammed frictionless systems, the model includes the constraint that the average number of contacting neighbors is six to satisfy the isostatic condition[24]. Since the model does not explicitly take into account positional disorder in the packing, it can only be applied to systems whose polydispersity exceeds 10% in radius [25] . In this paper, we improve the physical assumptions and the computational efficiency of the original granocentric model in order to capture the fluctuations in monodisperse 3D packings.

A granocentric model (first proposed in [19]) looks at the packing from the viewpoint of a single particle. Imagine that you take the place of a single particle in the bulk of the material. As you look out, your view of the system is blocked by the closest particles to you. A granocentric model describes the statistics of this first shell of neighbors without any consideration for other particles in the material. Here, we build on the success of the original granocentric model which we refer to as version 1.0 (v1), and create a version 2.0 (v2) that is additionally capable of describing monodisperse random packings. Model v2 captures key physical ingredients that go into packing particles: (1) filling the available space around a single particle with neighbors, (2) placing some of them in contact to ensure mechanical equilibrium and (3) approximating a volume for the local cell containing the central particle. Whereas model v1 treated these three steps sequentially and independently of one another to facilitate analytic solutions, the new model combines all the stages of the previous model into one search algorithm to determine the parameters self-consistently. This interdependence of the packing stages implies that model v2 represents physical reality more accurately. For example the determination of neighbors in stage (1) previously implied that they are all in contact with the central particle, while model v2 creates a cell in which some neighbors are in contact with the central particle and others are a given distance away. This and other improvements to model v1 introduce sufficient disorder into the model local cells to capture the distributions observed in monodisperse packings. The model parameters can be directly compared with measurable quantities describing the experimental packings.

This local model is tested against data from quasimonodisperse packings of poly(methyl methacry-late) (PMMA) particles, whose positions in 3D are imaged in the confocal microscope. The high resolution images allow for an estimation of coordination number from the geometric particle overlaps, while measurements of the occupied solid angle on each particle further test the underlying assumptions of the model. The agreement between the model and the experiment suggests that the model provides a valuable statistical tool for investigating packings in a wide range of applications. For example, we extend the model predictions to the local fluctuations in monodisperse packings ranging from random loose to random close packing fractions[11,26,27]. Within the granular statistical mechanics framework [3] , the model thus predicts the entropy[28] and the compactivity as a function of the global packing fraction and provides a way to map out a phase diagram of jammed matter[18,29]. More generally, the model parameters are derived from three global quantities: average coordination number, number of neighbors and the packing fraction, such that the fluctuations in any experimental packing with access to these quantities can be compared with the model predictions.

## 1.1 Granocentric model v2

The local packing structure around a given central particle is modeled by a stochastic process to fill the available solid angle with neighbors, each neighbor is assigned a distance from the central particle and a definition for the cell volume is then proposed. The details of this method appear in the online supplementary material.

Each local cell in the model starts with a center particle with a radius chosen at random from the input distribution (often a histogram of the experimental particle radii). Neighboring particles are sequentially added one at a time, with radii chosen from the same distribution as the central particle. Each neighbor is determined to be contacting with probability $p$, and otherwise placed a distance $\delta$ away from the central particle (chosen from a distribution with mean $\delta*$). We also add the physical constraint that at least $zmin$ neighboring particles are contacting to ensure local mechanical stability and exclude rattlers. Each neighboring particle then occupies a given solid angle, which is determined by the radii of both the neighbor and the central particle as well as the distance $\delta$ between them. Neighboring particles are added in this manner until the sum of the solid angle from all neighbors is greater than a threshold value, $\Omega*$. The last added neighbor is included into the neighbor shell only half the time, such that the physical interpretation of this threshold corresponds approximately to the average total solid angle, $\langle \Omega tot \rangle$, filled by the neighbors. The model therefore predicts the number of neighbors, the number of contacts and the total filled solid angle, $\Omega tot$, of a given central particle. Repeating this process yields the statistical fluctuations of these parameters within the packing.

In order to define a cell volume for a central particle we consider the volume contribution from each of its neighbors. Each neighbor occupies a solid angle that defines a cone (see Fig. 1) from the center of the central particle to the surface equidistant from both particles (defined by the navigation map30). Summing over these volumes from all the neighbors gives an estimate of the cell volume corresponding to the total filled solid angle $\Omega tot$. However, in real packings, cell volumes obtained by tessellation by definition occupy all the solid angle around each particle, i.e. 4pi. To account for the void space between neighboring particles when calculating cell volumes we partition the remaining unfilled solid angle (up to 4pi) between only the contacting neighbors and accordingly augment their cone volume contributions to the cell. The cell volume is then defined as the sum of the volumes of the augmented cones contributed by all neighbors. While the solid angles of the neighbors still sum to $\Omega tot$, the sum of the solid angles of their respective cones is 4pi. This also implies that the central particle's volume is exactly accounted for within the cell.

## 1.2 Determining the model parameters

The model relies on the knowledge of the particle size distribution and employs three adjustable control parameters: (1) the probability of contact with the central particle, $p$, (2) the solid angle threshold, $\Omega*$, and (3) the average surface-to-surface distance of non-contacting neighbors, $\delta *$. Using the algorithm described in Sec. 1.1 the model generates the probability distributions and therefore the average values for the number of neighbors, $n$, the number of contacts, $z$, the cell volumes, $V$, and the global packing fraction $\phi$ (approximated by the ratio of average particle volume to average local cell volume). In order to use the model to describe experimental packings we optimize the model control parameters to best match the experimental values of $\langle n \rangle$, $\langle z \rangle$ and the global density, $\phi$. A comparison of the experimental distributions of the local parameters with the model predictions serves as a test of the validity of the model and its assumptions. Moreover, we check whether the optimized control parameters ($p$, $\Omega*$, $\delta*$) correspond to their experimental counterparts.

Next, we describe the model in more detail and present the dependence of the model parameters on the experimental inputs. Although we are able to write exact equations (found in Sec. A of the online supplementary material) relating the model parameters to the output statistics, we find solving these equations is difficult. Therefore, we resort to efficient Monte Carlo simulations to create numerous local cells and calculate the parameters. This way we not only accurately obtain the mean quantities of interest, but also the entire distributions for number of neighbors, number of contacts, local cell volume, local packing fraction, and filled solid angle.

Numerous techniques exist to optimize model parameters. Here, we use a combination of reducing parameter space and the surface plots shown in Fig. 2, generated with an efficient Monte Carlo algorithm, to determine the optimal model control parameters. We begin by setting the control parameter, $p$, to the value $p = (\langle z \rangle - zmin)/(\langle n \rangle - zmin)$, where each cell must have at least $zmin$ contacting neighbors. Then, we create the surface plots in Fig. 2: the average number of neighbors (top) and the global packing fraction (bottom) for various values of $\Omega*$ and $\delta *$. Naively, one might generate each point with its own set of Monte Carlo simulations, but we use only one set of random numbers to generate the entire surface (as described Secs. B and C of the online supplementary material). The basic idea is to generate a databank of potential contacting and noncontacting neighbors. Next, for each pair of model parameters ($\delta*$, $\Omega*$) the non-contacting neighbors are pushed away from the surface so that the mean distance to the surface is $\delta*$ and we then determine which neighbors fill the available solid angle up to $\Omega*$. From the generated surface plots (Fig. 2), we select the thick black contour lines along which the average number of neighbors and global packing fraction match the experimental values ($\langle n \rangle = 14.4$, $\phi = 0.636$). This method identifies the model solution for $\Omega*$ and $\delta*$ as the point in parameter space where the two lines that satisfy experimental constraints cross. Finally, we compare the average number of contacts from the model to the average from the experimental data, adjust $p$ accordingly, and repeat the above process until all three parameters are determined within desired tolerances. Note that as $\delta*$ tends towards zero, the positional randomness decreases and leads to discrete integer solutions in that region of parameter space.

## 1.3 Comparison with the original granocentric model

Given that the granocentric model v1 is successful in describing polydisperse emulsion packings, we first demonstrate the ability of model v2 to describe the same experimental packings and compare the two methods. In Fig. 3 we show the confocal image of a randomly packed polydisperse emulsion and the corresponding reconstructed image of the droplet radii and centers, as described in[19,31]. Image analysis yields the number of neighbors $n$, contacts $z$ and the cell volume $V$ for each particle in the packing. Both the original granocentric model and the one presented here are equally good at describing polydisperse packings in terms of the previously measured distributions of $n$, $z$ and $V$, as shown in Fig. 3(a), (b), and (c). Nevertheless, disagreement between the original model and physical reality is apparent in the distribution of the total filled solid angle, $\Omega tot$, around the central particle. The model

v1 consistently overestimates $\Omega$tot compared to experimental data, as shown in Fig. 3(d), because it assumes that all neighbors are in contact with the central particle when filling the available solid angle $\Omega$max, which is a parameter in the model. Even though $\Omega$max seems to be an approximate upper bound for the total filled solid angle, it does not correspond to a measurable quantity in a real packing; it could be below 4pi due to shielding of neighbors or above 4pi because of overlapping solid angles. Instead, model v2 replaces $\Omega$max with the parameter $\Omega*$, which corresponds roughly to the experimentally measured average total solid angle filled by the particles, as shown in the figure.

In the model v1 some of the allocated neighbors are moved a distance $\delta$ away from the central particle to fit the global packing density. However, this process does not influence the number of neighbors since the neighbor selection step is independent of the step to create the cell volume. In a real packing, the further away the neighbors are the less solid angle they occupy and therefore more neighbors can be fit around a given particle. Therefore, model v2 includes an interdependence of all the parameters to satisfy this physical constraint. This improvement to the model leads to a much better agreement with the experimental distribution of total filled solid angle, as shown in Fig. 3(d). In conclusion, although both models capture the distributions of $n$, $z$ and $V$ in polydiserse packings, the values of the model parameters differ because they are inter-related in model v2 such that they have a stronger physical basis. age analysis yields the number of neighbors $n$, contacts $z$ and the cell volume $V$ for each particle in the packing. Both the original granocentric model and the one presented here are equally good at describing polydisperse packings in terms of the previously measured distributions of $n$, $z$ and $V$, as shown in Fig. 3(a), (b), and (c). Nevertheless, disagreement between the original model and physical reality is apparent in the distribution of the total filled solid angle, $\Omega$tot, around the central particle. The model v1 consistently overestimates $\Omega$tot compared to experimental data, as shown in Fig. 3(d), because it assumes that all neighbors are in contact with the central particle when filling the available solid an

## 2 Application to monodisperse packings

The granocentric model v1 uses the size distribution as the only source of randomness at the first stage of selecting neighboring particles, resulting in a delta function for the distribution of neighbors in the monodisperse case. Experimentally, this is not the case due to positional disorder. Indeed, monodisperse packings have been investigated by numerical simulations and experiments on particles ranging from ball bearings[7,8] to colloidal spheres[32]. The probability distribution of cell volumes is consistent between various monodisperse random close packings, as shown in Fig. 4. To gain access to all the local parameters in experimental random close packings (RCP), we study the jammed structure of fluorescently dyed athermal poly(methyl methacrylate) (PMMA) particles, shown in Fig. 5 (top left). The particles are sedimented under gravity and lubricated by a refractive index matched suspending medium that eliminates friction and renders the packing transparent for optical observation. We thus image a packing of over 6,000 particles, find each particle's location and size and reconstruct the original mage as shown in Fig. 5 (top right). To eliminate the small amount of polydispersity in the experiment, we only consider those particles whose radii are $\langle r \rangle = 1.65$ μ m $\pm$ 1%, which is the resolution of our particle finding algorithm. We are justified in taking such a subset of particles as we find this sampling does not bias the measurable global quantities such as the average coordination number, the average number of neighbors, or the global packing fraction from the original data set with all particles. The experimental packings are analysed to obtain local cell statistics (cell volume and nearest neighbors) by computing a Voronoi tessellation. Contacts are determined as those with a surfaceto-surface distance below a set resolution tolerance (0.04 times the particle radius). Since the distribution of Voronoi volumes is in good agreement with previously published monodisperse RCP results, as shown in Fig. 4, we next compare the experimental distributions of local parameters to those generated by the granocentric models.

Allowing each non-contacting neighbor in the model v2 to be assigned a surface-to-surface distance $\delta$ = $\delta*$ replaces the delta function predicted by model v1 by a broader distribution of neighbors that is closer to that observed in the experiment, as shown in Fig. 5(a). Moreover, this positional disorder also describes the distribution of filled solid angle (data not shown), as in the polydisperse case. On the other hand, the distributions of contacts are in good agreement between the two models, and accurately describe the data (Fig. 5(b)). This makes sense because the contact distribution is dominated by the probability $p$ of choosing contacts among neighbors, present in both models.

Since both granocentric models do not predict a smooth distribution of volumes seen in the experimental data in Fig. 4, further sources of randomness need to be introduced. A fixed value of $\delta = \delta*$ for the non-contacting neighbors leads to each pair of $n$ neighbors and $c$ contacts having the same volume and hence a discrete volume distribution. The ease of implementation of the models via Monte Carlo simulations allows us to add positional randomness by choosing delta from a probability distribution rather than a single value. The model allows for any choice of distribution, but we resort to a beta distribution with density $2(1-\delta/3\delta*)/3\delta*$ for $0 \leq \delta \leq 3 \delta*$ and zero otherwise. This distribution has mean $\delta*$ and does approximate the experimental distribution of surface-to-surface distance, while the optimized parameter $\delta*$ aligns beautifully with the experimental average, as shown in Fig. 5(c). Using the actual experimental distribution for $\delta$ is just as successful at reproducing the local fluctuations, therefore we use the minimal input needed to agree with experiments.

The additional randomness in the surface-to-surface distance broadens the neighbor distribution for model v2, making it a closer match to the experimental data (Fig. 5(a)). For the model v1, this additional positional randomness does not affect the neighbor nor the contact distributions but it does smooth out the volume distribution. However, this volume distribution is too broad compared to the experimental data. By contrast, model v2 predicts a smooth volume distribution in excellent agreement with the experimental data, as shown in Fig. 5(d).

In conclusion, model v2, with the chosen beta distribution, captures the distributions of the number of neighbor and contacts, as well as the filled solid angle and the local cell volume distribution in monodisperse packings. Although this is an minimal model, it accurately captures the fluctuations in real systems.

## 3 Phase space of jammed random packings

The granocentric model v2's success of describing the monodisperse packing at the random close packing (RCP) limit in Sec. 2 leads us to use this model to generate random configurations within a hypothetical phase space of jammed states. The way in which particles pack is influenced by many parameters, such as polydispersity, friction, rigidity, and the protocol by which the packing is created. Globally, these parameters change the density at which the particles pack and the average number of contacts between them, thus defining a phase space of jammed configurations accessible to real packings. For example, monodisperse particles are known to pack between random loose and random close packing densities depending on the friction coefficient [18,35] or the packing protocol [5,36] . No matter how a packing is generated experimentally, our granocentric model can be used to generate the fluctuations inside the packing using only global quantities as an input.

Here, the granocentric model is used to describe a wide range of monodisperse packings and to compare the results with existing theoretical approaches to jammed matter. In order for the model to work, the user inputs $\langle n \rangle$, $\langle z \rangle$ and $\phi$ for the packing of interest. We first generate the distribution of local cell volumes at the random loose packing (RLP) ( $n$ = 14, $\langle z \rangle$=4, $\phi \sim 0.53$) and at the RCP ( $n$ =14, $\langle z \rangle$ = 6, $\phi \sim 0.64$) limits for comparison, shown in Fig. 6(a). The loose configurations have larger cell volumes with a broader distribution than the RCP configurations. This decrease in the standard deviation between RLP and RCP has been observed in experimental packings [26,27,29,37] and indicates that the number of possible local configurations also decreases.

It is also interesting to relate the granocentric model to Edwards' statistical mechanics framework [3] for granular matter. Working under the hypothesis that packings occupying the same total volume have the same macroscopic properties, Edwards proposed to describe a randomly packed state using thermodynamic quantities that are analogous to those used in thermal systems: the system volume replaces the energy, and the compactivity replaces the temperature. Within this framework, a key quantity is the equivalent of the Gibbs entropy of the packing, which counts the number of microscopic configurations with a given total volume. While the Gibbs entropy is not accessible to the granocentric model since it requires the knowledge of the joint probability distribution of the volumes of all the cells in the packing, its approximation by the Boltzmann entropy can be readily calculated from

$$\int_0^\infty p(V)\log p(V)dV \quad (1)$$

where $p(V)$ denotes the probability density of a single cell in the packing. It is well-known that the Boltzmann entropy in Eq. (1) is a good approximation of the Gibbs entropy if correlations between the cells are weak[38] (in the standard equilibrium statistical framework this approximation becomes exact in the limit of an ideal gas), which is consistent with the local viewpoint taken in the granocentric model. Using this model to estimate $p(V)$ we can therefore map out the dependence of $S$ on the global quantities $\langle z \rangle$ and $\phi$, as shown in Fig. 6(b). In these calculations, we fixed $\langle n \rangle = 14$ as measured for the monodisperse RCP case because it does not change significantly between different packing densities[32]. The range of $\langle z \rangle$ is chosen to be between 4 and 6 because these values correspond to the isostatic condition in the limits of infinite and zero friction, respectively. The range of $\phi$ is chosen from below RLP at $\phi = 0.45$, since colloidal gels are known to jam at arbitrarily low values[39], to the RCP value of $\phi \sim 0.64$ to avoid crystallization. The granocentric model captures the general trend that is also seen in the literature: $S$ decreases with increasing packing density[17,29]. Moreover, $S$ increases with the average coordination number at a given density, consistent with the fact that there are more ways to choose 6 rather than 4 contacts out of the set of 14 neighbors.

Within Edwards' statistical mechanics framework[3], the entropy can be related to the compactivity, $\chi$, as

$$\chi^{-1} = \partial S / \partial v* \quad (2)$$

where $v = \langle V \rangle = V_p / \phi$ is the average cell volume and $V_p$ is the particle volume. Using formula (2) with $\langle n \rangle = 14$ fixed and $\langle z \rangle$ and $\phi$ ranging between 4 and 6 and 0.45 to 0.64, respectively, gives the result shown in Fig. 6(c) which indicates that the granocentric model captures our physical intuition about compactivity. States with lower density have higher compactivity, and therefore a greater ability to compactify further. Similarly states with fewer contacts have higher compactivity than states with the same density and more contacts.

Interestingly, measuring relative compactivity via Eq. (2) circumvents assuming a specific distribution for the volume but permits to test various distributions that have been proposed in the literature. For example, a popular model is the *k-gamma* distribution proposes in[16] which assumes that the volume fluctuations are described by

$$p(V)=k!/(k-1)!(V-V_m)^{k-1}/(v*-V_m)^k \exp\left(-k(V-V_m)/(v*-V_m)\right) \quad (3)$$

where $V_m$ characterizes the minimum volume (taken to be the volume of an FCC unit cell) and $k$ is the shape parameter. While our granocentric model generated RCP packing is well described by this gamma distribution, as are the experimental packings in this state, the granocentric generated distributions are less well fit as we approach the RLP limit (Fig. 6(a)). This is corroborated by the fact that fitting the data with a *k-gamma* distribution using either a maximum likelihood method or by extracting the parameters from the mean and the standard deviation of the data gives different values for the parameter $k$ (as apparent from the inset in Fig. 6(a)).

## 4 Conclusions

We have introduced a granocentric model capable of describing the fluctuations in the numbers of neighbors and contacts as well as the cell volumes and filled solid angle for experimental polydisperse and monodisperse random packings. This new version of the model is widely applicable as (1) it is capable of capturing the positional disorder as the dominant source of randomness in monodisperse packings, (2) the three control parameters are based on readily available experimental quantities and (3) it is optimized for efficient implementation.

As a test of this model, we have mapped out its predictions for a wide range of monodisperse packings with different densities and average coordination numbers corresponding to to our own random close packed PMMA packings and those encountered in the literature. Interestingly, the volume fluctuations predicted by the model are in good agreement with experimental data, indicating that the model captures the dominant physical features of granular materials. Furthermore, we make predictions of a plausible phase diagram of jammed matter according to a statistical mechanics framework, with which future experiments can be readily compared.

**Acknowledgments:** We thank Aleksandar Donev for giving us access to the numerical simulation data on monodisperse packings and Karen Daniels for useful discussions. J. B. holds a Career Award at the Scientific Interface from the Burroughs Wellcome Fund and was supported in part by New York University Materials Research Science and Engineering Center Award DMR-0820341 and a Career Award 0955021. E. V. was supported by NSF grant DMS07-08140 and ONR grant N00014-11-1-0345.

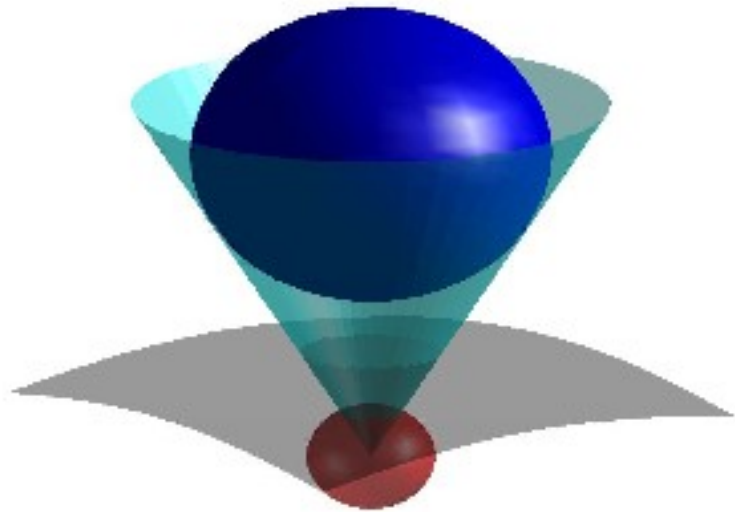

**Fig. 1** (color online) A visualization of the volume approximation for each neighbor in the granocentric model v2 described in Sec. 1.1. The cone represents the filled solid angle by the neighbor particle (large, blue online, sphere). The volume contribution of this neighbor is the volume of the cone between the center of the center particle (small, red online, sphere) and the hyperbolic sheet (grey surface) defining the navigation map between the surfaces of the two spheres.

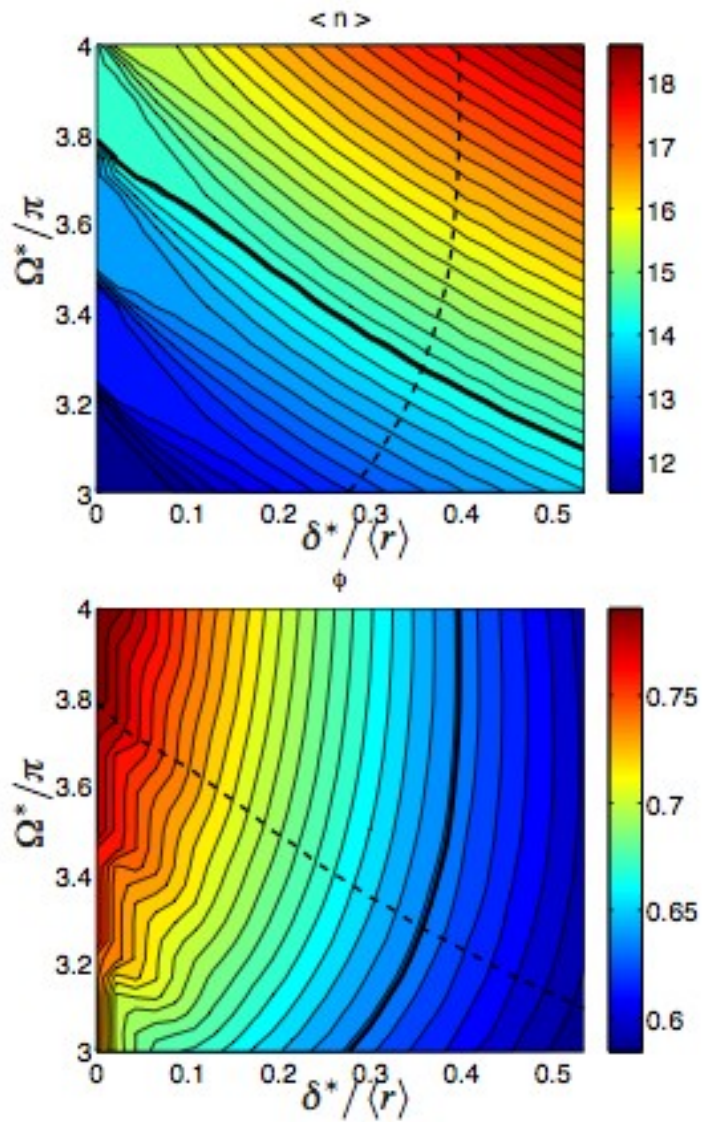

**Fig. 2** (color online) The average number of neighbors, $\langle n \rangle$, (top) and the global packing fraction, $\phi$, (bottom) generated by the granocentric model v2 as a function of the model control parameters $\Omega*$ and $\delta*$ (scaled by average particle radius) while keeping $p = 0.40$. In these generated surface plots the thick black contour lines indicate where the average number of neighbors (in the top plot) and global packing fraction (in the bottom plot) match the experimental values ( $n = 14.4$, $\phi = 0.636$). These curves also appears in the alternate plot as a dashed line.

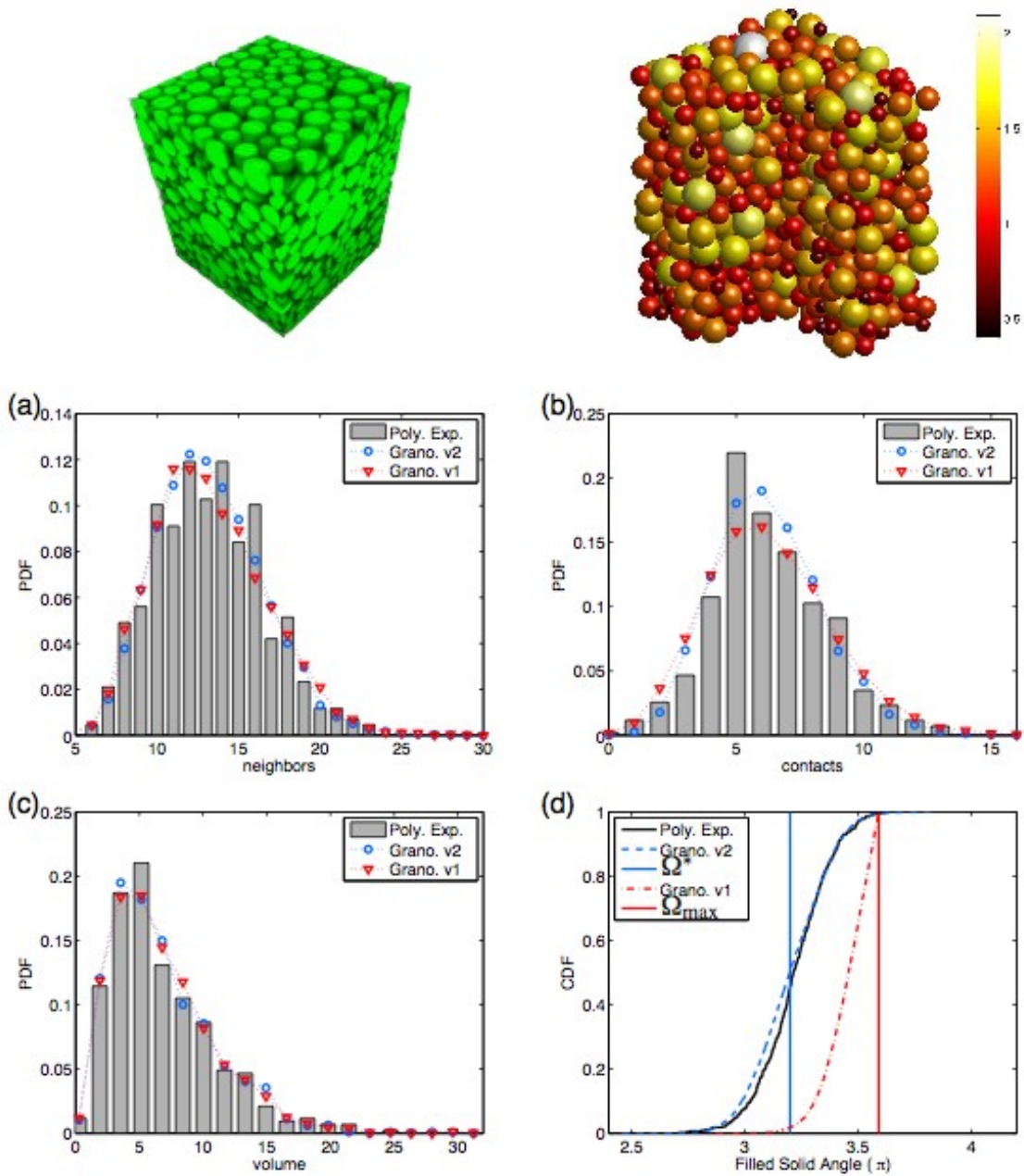

**Fig. 3** (top left) Confocal image and (top right) 3D reconstruction of polydisperse oil-in-water emulsion containing ≈ 1500 droplets in a box of volume $65 \times 65 \times 100 \mu m^3$. In the reconstruction, spheres are colored according to their radius, as indicated by the color bar, in units of average radius. Remaining figures compare experimentally obtained distributions to those created with the granocentric models v1 and v2: (a) probability a cell has $n$ neighbors, (b) $z$ contacts, (c) local cell volume $V$ in units of $\langle r \rangle^3$, and (d) cumulative distribution function for filled solid angle. Only a single value of $\delta = \delta^*$ (or $\delta = \hat{\delta}$ for v1) is used. For the new model, at least one contact is required to accommodate the allocation of remaining solid angle in the volume calculation.

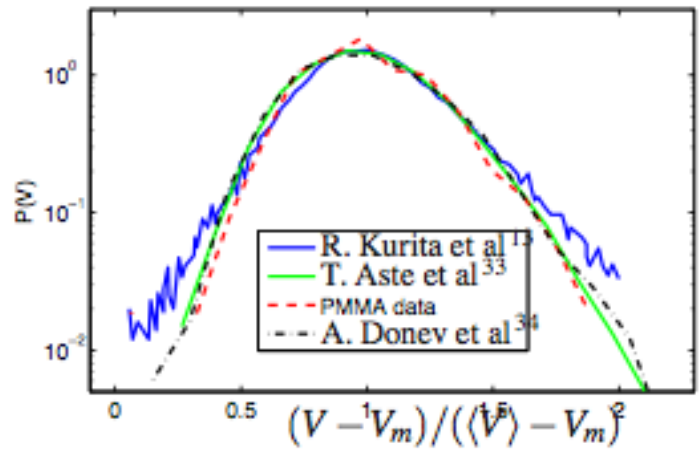

**Fig. 4** Distributions of the Voronoi cell volumes, $V$, plotted as a function of $(V - V_m)/(\langle V \rangle - V_m)$ for various monodisperse data sets indicated in the figure legend. Here, $\langle V \rangle$ is the average local cell volume obtained from each data set and $V_m$ characterized the minimum local cell volume, for the PMMA data it is the minimum volume of the data set as the theoretical minimum is unknown due to some polydispersity.

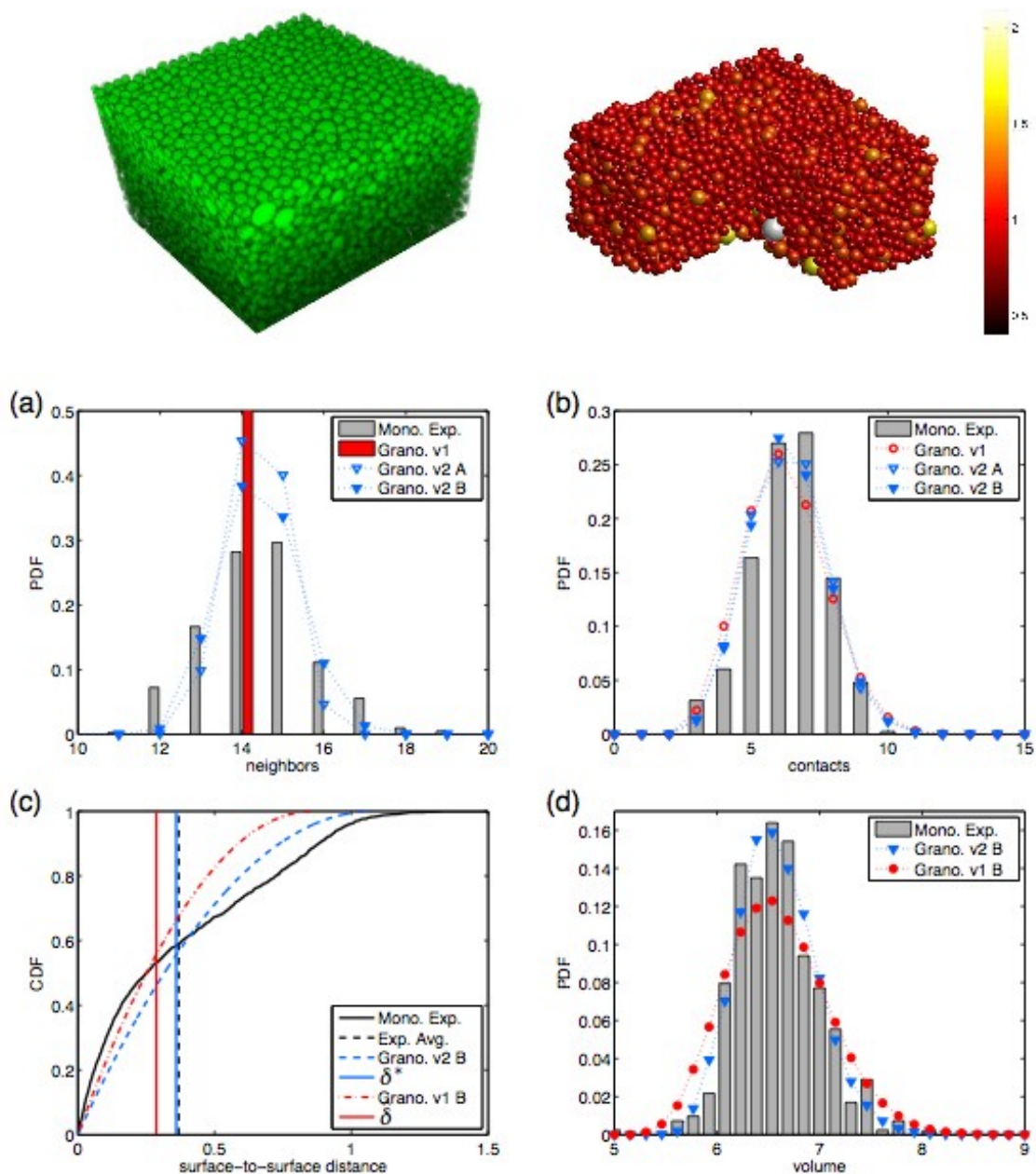

**Fig. 5** (top left) Confocal image and (top right) 3D reconstruction of monodisperse PMMA particles containing ≈ 6000 particles in a box of volume $70 \times 70 \times 35 \mu m^3$. In the reconstruction, spheres are colored according to their radius, as indicated by the color bar, in units of average radius. Remaining figures compare experimentally obtained distributions to those created with the granocentric models v1 and v2: (a) probability a cell has $n$ neighbors, (b) $z$ contacts, (c) cumulative distribution function for the surface-to-surface distance, $\delta$, in units of $\langle r \rangle$, and (d) probability distribution function for local cell volume, $V$, in units of $\langle r \rangle^3$. In granocentric models with "A" only one value of $\delta = \delta^*$ (or $\delta = \hat{\delta}$ for v1) is used, while in models with "B" the value of $\delta$ is chosen from the prescribed beta distribution. For all models, a minimum of 3 contacts are required to align with the experimental data.

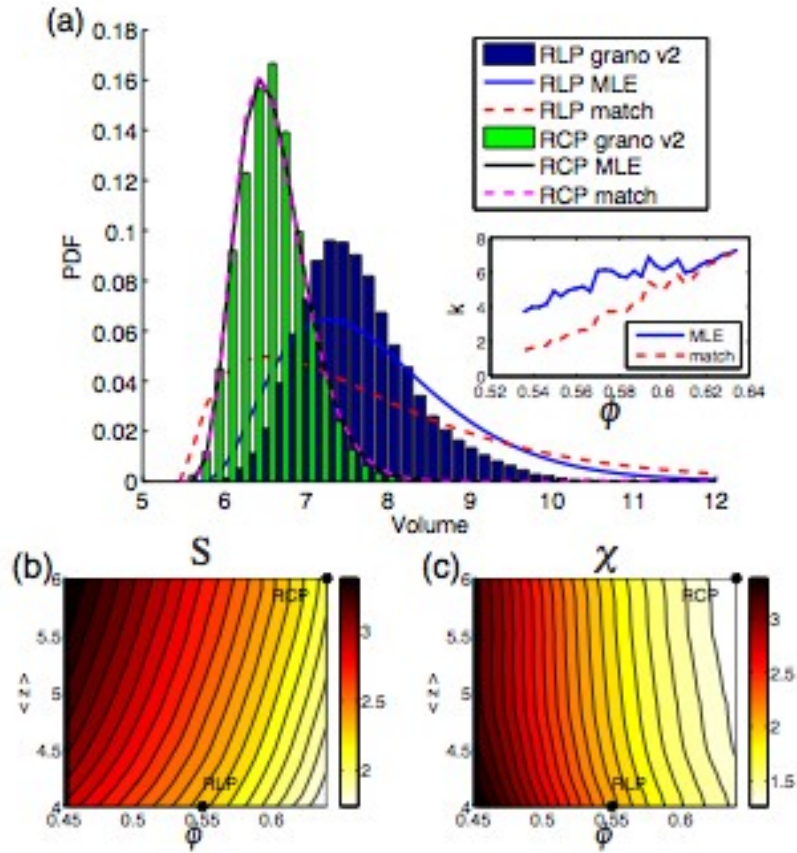

**Fig. 6** Using the granocentric model v2 with the prescribed beta distribution for δ we generate local cell volume statistics. (a) The distributions of cell volumes along with two different fittings of the *k-gamma* distribution, Eq. 3, at both the random loose packing (RLP) limit and the random close packing (RCP) limit. The scale and shape parameters of the *k-gamma* distribution are found using the maximum likelihood measurement (MLE) as well as by matching the mean and standard deviation of the distributions (match). The inset depicts the values of the shape parameter, *k*, over a range of packing densities between RLP and RCP. (b and c) We create packings over a spectrum of packing densities and average number of contacts while keeping the average number of neighbors constant at 14. From these model generated volumes we obtain (b) the Boltzmann entropy of the packing using Eq. (1) and compute (c) the compactivity, χ, using Eq. (2). Note that χ is given in units of $Vp = 4\pi/3 \approx 4.2$.